\documentclass[conference]{IEEEtran}
\IEEEoverridecommandlockouts
\usepackage{cite}
\usepackage{amsmath,amssymb,amsfonts}
\usepackage{algorithmic}
\usepackage{graphicx}
\usepackage{textcomp}
\usepackage{xcolor}
\def\BibTeX{{\rm B\kern-.05em{\sc i\kern-.025em b}\kern-.08em
    T\kern-.1667em\lower.7ex\hbox{E}\kern-.125emX}}
\begin{document}

\title{Truth Without Comprehension: A BlueSky Agenda for Steering the Fourth Mathematical Crisis
}

\newcommand{\tightauthor}[2]{%
  \begin{tabular}[t]{@{}c@{}}#1\\[-0.25em]#2\end{tabular}}

% \author{%
% \IEEEauthorblockN{Runlong Yu\IEEEauthorrefmark{1}, Xiaowei Jia\IEEEauthorrefmark{1}}%
% \textit{\IEEEauthorrefmark{1}University of Pittsburgh}\\
% \texttt{\{ruy59, xiaowei\}@pitt.edu}%
% }

\author{%
\IEEEauthorblockN{Runlong Yu}%
\textit{University of Alabama}\\
\texttt{ryu5@ua.edu}%
\and
\IEEEauthorblockN{Xiaowei Jia}%
\textit{University of Pittsburgh}\\
\texttt{xiaowei@pitt.edu}%
}

\maketitle

\begin{abstract}
Machine-generated proofs are poised to reach large-scale, human-unreadable artifacts. They foreshadow what we call the Fourth Mathematical Crisis. This crisis crystallizes around three fundamental tensions: trusting proofs that no human can inspect, understanding results that no one can fully read, and verifying systems that themselves resist verification.
As a minimal yet principled response, we propose the Human Understandability ($\mathsf{HU}$) meta-axiom, which requires that every proof admits at least one projection that is resource-bounded, divergence-measured, and acceptable to a verifier. Confronting these questions opens a timely research agenda and points toward new directions in scalable reasoning, interpretable inference, and epistemic trust for the era of machine-scale mathematics.
\end{abstract}

%
%\begin{IEEEkeywords}
%component, formatting, style, styling, insert.
%\end{IEEEkeywords}

%With the exponential growth of computing power, the proliferation of formalized mathematical data, and radical advances in artificial intelligence (AI) methods for knowledge representation and automated reasoning, mathematical discovery has entered an era dominated by ``human-unreadable'' proofs. A landmark event occurred in 1976, when Appel and Haken proved the Four-Color Theorem by exhaustively verifying over a billion configurations using computational algorithms, since no mathematician could manually examine the argument step by step. Initially, the mathematical community expressed two core concerns: verifiability (Could hidden bugs exist within the code?) and intelligibility (Does a truth still qualify as mathematics if it cannot be ``seen'' or fully understood?). Since that pivotal moment, the traditional criterion that a mathematical proof should be personally comprehensible started to decouple from the community’s acceptance of mathematical truths. This divergence laid the foundation for the present-day \textbf{Fourth Mathematical Crisis}. Today, as machines generate proofs whose scale and complexity surpass the capacity of human lifetime verification, the implicit axiom ``mathematical truth $\equiv$ a form of argument entirely comprehensible to humans'' ceases to hold.

\setlength{\skip\footins}{5pt}

\section{What challenges our current assumptions?}
With the exponential growth in computing power, the expansion of large formalized mathematical corpora, and rapid advances in artificial intelligence (AI) techniques for knowledge representation and automated reasoning, mathematical discovery is undergoing a profound transformation. Terence Tao captures the spirit of this shift, suggesting AI will soon serve as a powerful co-pilot for mathematicians~\cite{Tao2024SciAm}. Recent breakthroughs highlight the accelerating pace of this evolution. Early neural theorem provers, such as \emph{GPT-f}, achieved around 29\% Pass@8 accuracy on the Lean \textsc{miniF2F} benchmark, often producing proofs involving hundreds of tactic steps~\cite{Polu2020GPTf,Zheng2022MiniF2F}. Subsequent neuro–symbolic systems have improved this to approximately 42\% Pass@32~\cite{Lample2022HTPS}. In geometry, \emph{AlphaGeometry} solved 25 of 30 problems in the \textsc{IMO–AG–30} benchmark, after training on roughly $10^8$ synthetic theorems~\cite{Trinh2024AlphaGeometry}. Most recently, DeepMind's \emph{Gemini} model in ``Deep Think'' mode achieved a gold medal under the official conditions of the IMO 2025 contest, marking a historic milestone for end-to-end natural-language mathematical reasoning~\cite{DeepMind2025GeminiIMO}.

% With the exponential growth of computing power, the expansion of large formalized mathematical corpora, and rapid advances in artificial intelligence~(AI) techniques for knowledge representation and automated reasoning, mathematical discovery is undergoing a profound transformation. Terence Tao captures the spirit of this shift, noting that AI may soon become a great co-pilot for mathematicians~\cite{Tao2024SciAm}. Recent breakthroughs underscore both the scale and the accelerating pace of this transition.
% Early neural theorem provers such as \emph{GPT\textnormal{-}f} achieved approximately 29\% Pass@8 accuracy on the Lean \textsc{miniF2F} benchmark, often generating proofs that spanned hundreds of tactic steps~\cite{Polu2020GPTf,Zheng2022MiniF2F}. More recent neuro–symbolic systems have pushed this performance to roughly 42\% Pass@32~\cite{Lample2022HTPS}. In geometry, \emph{AlphaGeometry} solved 25 out of 30 problems in the \textsc{IMO–AG–30} benchmark after training on a synthetic corpus of roughly $10^8$ theorems, achieving performance comparable to recent International Mathematical Olympiad gold medalists~\cite{Trinh2024AlphaGeometry}.

An increasing body of evidence suggests that we are entering an era dominated by ``human-unreadable'' proofs. A landmark event foreshadowing this trend occurred much earlier, in 1976, when Kenneth Appel and Wolfgang Haken settled the Four-Color Theorem. They reduced the problem to 1,936 unavoidable configurations and used thousands of hours of then–state-of-the-art computation to verify each case~\cite{AppelHaken1977a,AppelHaken1977b,Tucker1977FourColor}. While the proof was logically sound and formally checkable, its sheer scale rendered it nearly inaccessible to any individual mathematician, marking one of the first instances where verification and understanding began to diverge. At the time, mathematicians voiced two fundamental concerns: Could hidden bugs lurk within the computational code? Does a mathematical truth remain meaningful if it cannot be directly ``seen'' or fully understood by humans? From that pivotal moment, the traditional criterion that mathematical proofs should be fully comprehensible to individual mathematicians began to decouple from the community’s acceptance of mathematical truth, laying the foundation for what we term the \textbf{Fourth Mathematical Crisis}~\cite{MacKenzie2001Mechanizing,Buzzard2021Proof}. 
Looking ahead, AI systems are rapidly approaching the point where they generate proofs whose scale and complexity will surpass what any human could verify in a lifetime. As a result, the once-implicit axiom that ``mathematical truth~$\equiv$~a form of argument entirely comprehensible to humans'' is rapidly losing force. We contend that this emerging crisis foregrounds three fundamental questions that challenge the assumptions of the discipline:

\begin{enumerate}
	\item When a theorem is declared proven because an algorithm claims to have exhaustively searched an astronomically large space, yet neither full human inspection nor any feasible sampling strategy can establish meaningful confidence in the result, how are we to trust such a proof? 
	\item When an AI-generated proof spans logical steps far beyond what any human could read or trace in a lifetime, what qualifies as meaningful understanding of the result? 
	\item When the computational systems and algorithms that generate or check such proofs become too large, layered, or opaque for full human audit, how do we escape the recursive dependency on unverifiable verifiers?
\end{enumerate}

Confronted with these challenges, we critically evaluate humanity’s evolving role within an AI-driven proof ecosystem. If this crisis remains unaddressed, human involvement risks devolving into symbolic oversight, mathematicians merely rubber-stamping lists of machine-derived proofs they can no longer meaningfully comprehend. In this scenario, mathematics would persist only as an ``extraterrestrial artifact,'' formally correct and useful to machines, yet fundamentally opaque and inaccessible to the human minds that originally conceived it.

%\textit{(Challenges the assumption: Mathematical truth should remain within the bounds of human-time-limited verifiability.)}
%\textit{(Challenges the assumption: Every mathematical truth can ultimately be rendered in a form comprehensible to humans.)}
%\textit{(Challenges the assumption: The correctness of our verification tools can always be transparently established through human inspection.)}

\section{What should we ponder about it?}

\textbf{\emph{Q1:}} Modern combinatorics offers striking examples of proofs that rely on exhaustive computer searches: finite in principle, yet far beyond the reach of human verification in practice. A prominent case is the 2016 solution to the Boolean Pythagorean Triples problem, whose SAT solver produced an unsatisfiable certificate of roughly 200~terabytes; replaying and checking this trace still requires about two days on a modern computing cluster~\cite{Heule2016BPT}. Similarly, the proof that the Schur number $S(5)=160$ yielded a certificate of 2~petabytes of deletion resolution asymmetric tautology (DRAT) proof data~\cite{Heule2018AAAI}. Even its compressed version ($\approx$~70 GB) remains far beyond the limits of a meaningful human audit.

In such cases, the core of the proof lies in the exhaustive verification of a vast, though finite, collection of computationally generated objects, such as graphs, SAT clause cubes, or colorings. Despite being formally finite, these sets are so astronomically large that human inspection is effectively impossible, and even partial verification becomes impractical. Attempts to build confidence through random sampling are likewise problematic: even billions of samples represent only a negligible fraction of the total space, offering minimal statistical assurance. Sometimes, validating even a single element from these large sets can be computationally intensive, often requiring specialized solvers or high-performance computing resources inaccessible to most mathematicians or auditors.

Mathematicians are increasingly compelled to trust the correctness of such results not through direct scrutiny. This marks a profound epistemic shift: trust in mathematical truth no longer stems solely from human-inspectable arguments, but hinges on the correctness of the computational systems and algorithms that produce or verify the results (see~\textit{Q3}).

\textbf{\emph{Q2:}}  While \textit{Q1} faces the challenge of verifying vast sets of computationally generated objects, \textit{Q2} confronts a different kind of complexity: the immense length and depth of single, linear chains of reasoning. 
Take Thomas Hales’ formal verification of the Kepler Conjecture in the \textsc{Flyspeck} project: an informal argument that fits into a few hundred printed pages explodes into roughly half a million lines of HOL Light code, expanding to hundreds of millions of primitive inference steps when checked by the kernel~\cite{Hales2017Flyspeck}.  
An even larger undertaking is the ongoing formalization of the classification of finite simple groups, which is expected to take billions of such steps when completed~\cite{Gonthier2013OddOrder,Gonthier2020CFSGSlides}.
At a generous human reading pace of about $10^{3}$ inference steps per hour, a single mathematician would require centuries to read through the entire proof once, let alone fully comprehend or critique it.

Historically, lengthy proofs are not unprecedented. Gentzen’s 1936 consistency proof already employed transfinite induction up to the ordinal~$\varepsilon_0$~\cite{Gentzen1936}.  The printed proof of the classification of finite simple groups (CFSG) itself is on the order of ten thousand journal pages~\cite{GorensteinLyonsSolomon1994}.  What distinguishes modern AI-generated proofs from these classical ``monster proofs'' is their apparent resistance to compression. Classical arguments, though enormous, usually admit an intuitive hierarchy of lemmas and key insights that mathematicians can summarise. In contrast, automated theorem-proving pipelines (especially those built around large language models and proof assistants such as Lean or Isabelle) often generate proofs with little discernible structure. Any attempt to prune or summarise these traces may break machine verification, as each low-level inference step can be logically indispensable. The foundational speed-up results in proof theory~\cite{Godel1936} show that some true statements do not admit substantially shorter proofs within a given system. It is plausible that AI provers are producing such minimally compressible proofs.

Thus, \textit{Q2} forces us to revisit what ``understanding'' means in mathematics. Traditionally, mathematicians have sought not only correctness but also intuitive insight and transferable concepts. When a proof grows beyond lifetime-scale readability, however, we face an uneasy trade-off: accept machine-checked correctness without genuine comprehension, or demand summaries that may omit inference steps essential to formal validity. Several strands in proof theory aim to bridge this gap.  Kreisel’s proof mining programme and Kohlenbach’s subsequent functional interpretations extract quantitative and conceptual content from otherwise unwieldy proofs~\cite{Kreisel1959Proofs,Kohlenbach2008BK}.  More recently, homotopy type theory (HoTT) offers higher-level encodings whose equivalence principles can compress whole families of arguments into a single synthetic path~\cite{HoTTBook2013}.  Yet these techniques require the original proof to exhibit exploitable regularities; when machine-generated traces lack any discernible structure, such extraction may fail entirely.

\textbf{\emph{Q3:}} The concerns raised by \textit{Q1} and \textit{Q2} naturally lead to an even deeper question: Can we confidently trust the computational systems and algorithms that generate or verify these proofs? Modern proof-assistant tools consist of tens of thousands of lines of sophisticated, optimized code, continuously developed by numerous contributors~\cite{deMoura2021Lean,CoqTeam2024}. Expecting mathematicians (or dedicated reviewers) to thoroughly audit and verify such massive codebases is neither practical nor realistic. Yet even a thorough audit of the software's logic would not fully resolve the issue. The correctness of these verification programs rests upon deeper, often invisible layers: compilers translating verification logic into machine-executable instructions~\cite{Leroy2009CompCert}, operating systems managing hardware interactions~\cite{Klein2009seL4}, and hardware microarchitectures implementing opaque layers of microcode, firmware, and speculative execution~\cite{Kocher2020Spectre}. 

Historical experience underscores the fragility of each link in the trust chain. 
In the early 2010s, for instance, a subtle kernel bug within the \textsc{Coq} proof assistant briefly allowed incorrect theorems to be certified as valid~\cite{CoqBug2015}.  
Hardware offers no absolute refuge: the 1994 Pentium \textsc{FDIV} flaw corrupted floating-point arithmetic at the silicon level~\cite{Intel1995FDIV}, and the Spectre/Meltdown class of speculative-execution attacks exposed covert channels deep inside modern CPUs~\cite{Kocher2020Spectre}.  
Ken Thompson’s classic ``Trusting Trust'' lecture famously showed that even a verified program can be subverted by a compromised compiler~\cite{Thompson1984Trust}.  
Formally verified artifacts such as the CompCert C compiler and the seL4 microkernel mark real progress~\cite{Leroy2009CompCert,Klein2009seL4}, yet they still rely on linkers, firmware, and physical hardware that escape rigorous audit.  
As proof-checking experiments migrate to GPU or FPGA accelerators, the scope for opaque faults may only grow.

% Historical experience underscores the real risk posed by each link in this complex trust chain. In 2011, for instance, a subtle kernel bug within the Coq proof assistant briefly allowed incorrect theorems to be certified as valid. Hardware-level vulnerabilities, like Intel's notorious \textsc{FDIV} arithmetic flaw or more recent speculative-execution exploits, further illustrate how even the fundamental arithmetic operations embedded in silicon cannot be fully guaranteed. Ken Thompson's classic ``trusting trust'' lecture vividly demonstrated that compilers themselves can invisibly introduce malicious code, perpetuating a cycle of contamination that undermines even formally verified software built upon them. While celebrated verification efforts such as the formally verified CompCert C compiler and the seL4 microkernel represent meaningful steps toward more trustworthy foundations, each still depends upon lower-level components (such as linkers, firmware, and ultimately physical hardware) that largely escape rigorous auditing. Furthermore, as proof-verification increasingly moves toward GPU acceleration or custom hardware solutions, the potential for opacity and hidden faults grows exponentially.

Thus, \textit{Q3} highlights a profound epistemic challenge: faced with proofs too extensive to check directly (\textit{Q1}) and too deep to meaningfully comprehend (\textit{Q2}), we are compelled to rely on automated verifiers, only to confront a recursive dependence on increasingly complex and unverifiable systems. Trusting the output entails trusting its generator, and so on. This dilemma is not resolved by incremental fixes or patching known vulnerabilities. Until deeper trust mechanisms are established, our acceptance of machine-generated mathematical proofs remains precariously dependent on software and hardware whose reliability we cannot fully examine or completely guarantee.

\textbf{\emph{Summary.}} Each of the first three mathematical crises emerged when a new concept profoundly challenged existing logical frameworks: the discovery of irrational numbers shattered the Pythagorean belief in universal commensurability; the rigorous formalization of limits resolved longstanding paradoxes involving infinitesimals; and Gödel’s incompleteness theorems exposed fundamental limits inherent in formal proof systems themselves. The \textbf{Fourth Mathematical Crisis}, however, differs crucially from its predecessors: it is driven not by the introduction of unfamiliar objects or axioms, but by the unprecedented scale and opacity of computer-generated proofs. Our foundational assumptions, that mathematical proofs are human-surveyable, that correctness inherently includes comprehensibility, and that the tools used for verification are transparently reliable, begin to fracture under modern conditions. Nonetheless, this crisis remains fundamentally mathematical, striking at the discipline’s epistemic core by raising the questions: \emph{What constitutes a valid proof? Who or what possesses the authority to certify mathematical truth?} Just as past mathematical crises gave rise to new notions of number, limit, and formal systems, addressing this fourth crisis will demand the introduction of new foundational axioms, though what they might be remains unclear. 

% This effort aims to re-anchor the concept of mathematical truth in an era increasingly shaped by forms of reasoning that lie beyond direct human inspection or full comprehension.

\section{What is the BlueSky idea?}

Resolving the Fourth Mathematical Crisis may seem like a distant goal. It requires a sustained, interdisciplinary effort among mathematicians, computer scientists, data scientists, and proof assistant developers. At its heart lies a philosophical divide. On one side stands the skeptical traditionalist, who asserts that a statement cannot be considered a proof if no human mind can understand its internal logic. For this view, mathematics is defined not only by formal correctness but also by the pursuit of insight. A 200-terabyte proof certificate offers no more explanatory value than a black-box oracle and may even hide subtle encoding errors beneath its surface. On the other side is the accelerationist, who is open to letting the machines lead. From this perspective, history often favors technology over understanding: Newtonian calculus came before $\varepsilon$–$\delta$ rigor by over a century, and numerical simulations in fluid dynamics anticipated formal turbulence theory by decades. Given a large enough collection of formalized theorems, they argue, new abstractions and intuitions may simply emerge.

This paper proposes a BlueSky agenda that seeks a principled middle path. Machine-scale proofs can remain mathematically meaningful, but only when interpreted relative to a new layer of the meta-axiom that governs how such proofs are stored, summarized, and re-verified. We introduce this layer as the	$\mathsf{HU}$ (\textbf{Human Understandability}) meta-axiom system.

Why ``meta'' axioms? Traditional axiom systems (e.g., ZFC, Peano arithmetic, type theory) operate within the object language of mathematics, specifying properties of sets, numbers, terms, and propositions. The $\mathsf{HU}$ meta-axiom, by contrast, operates at a higher level: it concerns the artifacts that certify mathematical propositions (proof traces, version histories, semantic compression certificates, and verification workflows). Thus, $\mathsf{HU}$ does not define new mathematical objects; it constrains the epistemic infrastructure that renders those objects intelligible and trustworthy to humans.

\smallskip
\noindent\textbf{Notation.}
\begin{itemize}
  \item $\mathcal{P}$ — set of complete proof artifacts (machine traces, formal scripts, certificates);
  \item $\mathcal{A}$ — community of ideal verifiers (humans, proof-assistant kernels, ZK checkers, specialised hardware);
  \item $D : \mathcal{P}\times\mathcal{P} \to \mathbb{R}_{\ge 0}$ — semantic–divergence metric;
  \item $\Phi_{c,d} : \mathcal{P} \to \mathcal{P}$ — approximation operator such that
        \[
           \operatorname{Cost}\bigl(\Phi_{c,d}(P)\bigr)\le c,
           \quad
           D\bigl(P,\Phi_{c,d}(P)\bigr)\le d
           \quad(\forall P\in\mathcal{P});
        \]
  \item $\boldsymbol{\beta},\boldsymbol{\delta} : \mathcal{A} \to \mathbb{R}_{>0}$ — per‑verifier resource budget and maximum tolerated divergence.
\end{itemize}

\smallskip
\noindent\textbf{Meta‑axiom \(\mathsf{HU}\).}
For every verifier \(a\in\mathcal{A}\) and every proof \(P\in\mathcal{P}\), there exist parameters
\[
(c,d)\in C(a)\;:=\;\bigl\{(c',d')\mid c'\le\boldsymbol{\beta}(a),\; d'\le\boldsymbol{\delta}(a)\bigr\}
\]
such that \vspace{-0.2cm}
\[
\fbox{\(
    \mathrm{Verify}\bigl(a,\Phi_{c,d}(P)\bigr)=\texttt{true}
  \)}
\]
$\text{and }
\operatorname{Cost} \bigl(\Phi_{c,d}(P)\bigr)\le c,\; D \bigl(P,\,\Phi_{c,d}(P)\bigr)\le d \text{ by definition.}$
    % \operatorname{Cost}\le c,\; D\le d 
    
\smallskip
\noindent\textbf{Reconstruction protocol.}
Let \(\Psi(t,\cdot):\mathcal{P}\to\mathcal{P}\) regenerate an artifact using the tool chain available at time \(t \ge 0\), where \(t\) is an abstract timestamp identifying a snapshot of the tool chain. For every \(a\in\mathcal{A},\,P\in\mathcal{P},\,t\ge 0\) there exist \((c,d)\in C(a)\) such that \vspace{-0.1cm}
\[
\fbox{\(
    \mathrm{Verify}\bigl(a,\Phi_{c,d} \bigl(\Psi(t,P)\bigr)\bigr)=\texttt{true}
  \)}
\]
with $\operatorname{Cost} \bigl(\Phi_{c,d}\bigl(\Psi(t,P)\bigr)\bigr) \le c, \; D \bigl(P,\,\Phi_{c,d} \bigl(\Psi(t,P)\bigr)\bigr) \le d$ guaranteed by the properties of \(\Phi_{c,d}\).

Thus, after any tool‑chain update, each verifier still receives an
instance whose cost stays within its budget and whose semantic
divergence never exceeds its declared tolerance~\(\boldsymbol{\delta}(a)\).

Traditionally, a proof is a static, monolithic artifact (every reader inspects in full). The $\mathsf{HU}$ axiom reframes proofs as malleable objects, equipped with a family of approximations $\Phi_{c,d}$ tailored to the verifier’s constraints. Here, $(c,d)$ quantifies the resource cost (e.g., time, memory, cognitive load) and the semantic divergence from the original. A verifier $a$ declares a tolerance box $C(a)$, and a proof is epistemically acceptable only if some $\Phi_{c,d}(P)$ falls within this box.

This makes verifiability adaptive. A large proof may be trusted via many lossy projections, each tailored to a verifier’s cognitive or computational limits. Consider the 200-terabyte SAT certificate for the Boolean Pythagorean Triples problem: for a verifier facing \textit{Q1}, $\Phi_{c,d}$ extracts a verifiable fragment under cost budget $\beta(a)$ and replaces the remainder with cryptographic hashes. For \textit{Q2}, involving long reasoning chains, a human may request a coarse projection summarizing the core contradiction, while delegating billions of low-level steps to machine verifiers. Under \textit{Q3}, reconstruction is handled by applying $\Psi(t,P)$ when the proof toolchain evolves; if the verifier still finds some $(c,d) \in C(a)$, the proof remains valid.

Rather than attempt to prove the $\mathsf{HU}$ meta-axiom, we outline a concrete implementation sketch to illustrate how such a human-understandability contract could function in practice. Our goal is to show that the proposal is technically grounded, epistemically meaningful, and realizable in principle. In a minimal setup, a complete proof $P$ could be represented as a Lean script hosted in a public Git repository~\cite{deMoura2021Lean}. As proofs grow in size and depth, their verification traces might be partitioned into fragments small enough to fit within a verifier’s declared resource budget~$\boldsymbol{\beta}(a)$ and streamed interactively, allowing each verifier, whether human or machine, to process only what it can afford.
To control semantic complexity, we may introduce a mechanism of semantic folding, in which deep subtrees of the proof directed acyclic graph (DAG), beyond a tunable threshold $\varepsilon$, are replaced with cryptographic hashes. This defines a family of approximations $\Phi_{\varepsilon}(P)$, each representing a compressed, resource-bounded projection of the original proof. A verifier would select an appropriate $\varepsilon$ such that the resulting approximation remains within their semantic tolerance~$\boldsymbol{\delta}(a)$. Although each verifier sees only a fragment of the full artifact, the global hash-chain preserves semantic integrity across the omitted parts~\cite{Chacon2014ProGit}.
As tools evolve over~time, the underlying environment used to build or verify $P$ may shift. In such cases, a continuous integration (CI) script could invoke a regeneration protocol $\Psi(t,P)$ to reproduce the artifact under the current toolchain. One would then seek a pair $(c,d)$ within the verifier’s tolerance box $C(a)$, such that the approximation $\Phi_{c,d}(\Psi(t,P))$ remains acceptable. If no such projection is found, the update is rejected. In practice, Lean minor releases are conservative, so either $D\bigl(P,\Psi(t,P)\bigr)=0$ or an admissible approximation $\Phi_{c,d}(\Psi(t,P))$ is quickly located; major version transitions, however, may still require semi-automated porting.

All steps in this process are fully Turing-computable and interpretable within standard foundations. The $\mathsf{HU}$ meta-axiom introduces no new object-level primitives, only constraints on how proofs should be structured and maintained to remain interpretable by bounded agents. Conversely, ZFC extended with $\neg\mathsf{HU}$ is just as consistent: one relinquishes any requirement that proofs be human-accessible. The novelty of $\mathsf{HU}$ lies in its infrastructure: it offers an epistemic layer to preserve a human-facing boundary in an era of machine-scale mathematics.

However, the working premise behind $\mathsf{HU}$ (there exists at least one projection simultaneously within human cognitive limits and sufficiently semantically faithful) may fail for some classes of proofs. In such cases, \textbf{mathematics can be divided into \emph{human-understandable} and \emph{human-non-understandable} mathematics}, a distinction that significantly departs from traditional conceptions of the discipline.

\section{Challenges and what will success look like?}

We turn to the key challenges that arise in realizing the $\mathsf{HU}$ meta-axiom.
First, mathematicians are asked to undergo an epistemic shift: to accept that a proof may be verifiable without ever being fully understandable. Verifiers engage only with resource-bounded projections, partial glimpses of a structure whose totality may lie forever beyond human cognition. The psychological cost is profound: for centuries, mathematical truth has been intertwined with the potential for full human comprehension. To relinquish that hope is to confront the unsettling reality that some theorems, though provably true, may remain cognitively inaccessible not just to individuals, but to humanity as a whole.
Second, the framework quantifies trust through parameters $(c, d, \boldsymbol{\beta}, \boldsymbol{\delta})$, yet lacks a canonical measure of semantic divergence. What counts as an acceptable loss of meaning varies across disciplines, cultures, and even individuals, making $\boldsymbol{\delta}$ not just a technical threshold but a site of epistemic negotiation. 
Third, once proofs are sliced into cost-aware views, computational capital becomes a gatekeeper: high-fidelity verification may become the privilege of those with GPUs and parallel clusters, while others are left with coarse approximations. Fourth, the entire pipeline (from $\Phi$ to $\Psi$) rests on a fragile software–hardware stack. Any break in that stack, whether due to toolchain drift, bit-rot, or post-quantum vulnerabilities, can silently erode trust and revive the recursive-verifier dilemma posed by \textit{Q3}.

Despite these obstacles, $\mathsf{HU}$ offers a powerful new alignment between formal rigor and human insight. Like human Go players who gained new intuition by studying AI gameplay, mathematicians may find their own understanding enriched by engaging with machine-scale reasoning under principled constraints. A mature ecosystem would make large proofs navigable, trusted, and socially accountable. Proofs could be decomposed into low-cost, low-divergence projections, each annotated with key lemmas and conceptual structure to support human reasoning. Every slice would carry verifiable provenance; toolchain upgrades would trigger automatic re-validation. Divergence budgets~$\boldsymbol{\delta}(a)$ would be governed transparently, enabling verifiers to trade accuracy for efficiency while leaving a public trace. Communities could audit these trade-offs or fund deeper projections for theorems of special importance. In such a future, mathematics would no longer require universal comprehension to be meaningful. It would remain deeply human, grounded in shared mechanisms for trust, intuition, and interpretation.

\section{Why should data scientists care?}

Machine-generated proofs are becoming some of the largest and most intricate data artifacts ever produced, containing billions of inference steps, petabytes of symbolic traces, and layered provenance from software to hardware. They are evolving knowledge graphs that encode the structural regularities of mathematics. Mining such proofs could redefine key areas of data mining, such as representation learning, graph compression, and trustworthy AI. The proof deluge offers a new testbed for foundational advances. Scientifically, we gain access to inaccessible knowledge, including latent proof structures, emergent conjectures, and formally minimal explanations. Societally, it prepares us for domains where autonomous systems already exceed human audit.

\bibliographystyle{IEEEtran}
	\bibliography{mybibliography}

\begin{thebibliography}{10}
\providecommand{\url}[1]{#1}
\csname url@samestyle\endcsname
\providecommand{\newblock}{\relax}
\providecommand{\bibinfo}[2]{#2}
\providecommand{\BIBentrySTDinterwordspacing}{\spaceskip=0pt\relax}
\providecommand{\BIBentryALTinterwordstretchfactor}{4}
\providecommand{\BIBentryALTinterwordspacing}{\spaceskip=\fontdimen2\font plus
\BIBentryALTinterwordstretchfactor\fontdimen3\font minus
  \fontdimen4\font\relax}
\providecommand{\BIBforeignlanguage}[2]{{%
\expandafter\ifx\csname l@#1\endcsname\relax
\typeout{** WARNING: IEEEtran.bst: No hyphenation pattern has been}%
\typeout{** loaded for the language `#1'. Using the pattern for}%
\typeout{** the default language instead.}%
\else
\language=\csname l@#1\endcsname
\fi
#2}}
\providecommand{\BIBdecl}{\relax}
\BIBdecl

\bibitem{Tao2024SciAm}
C.~Dr{\"o}sser, ``Ai will become mathematicians' `co-pilot','' \emph{Scientific
  American}, June 2024.

\bibitem{Polu2020GPTf}
\BIBentryALTinterwordspacing
S.~Polu and I.~Sutskever, ``Generative language modeling for automated theorem
  proving,'' \emph{CoRR}, vol. abs/2009.03393, 2020. [Online]. Available:
  \url{https://arxiv.org/abs/2009.03393}
\BIBentrySTDinterwordspacing

\bibitem{Zheng2022MiniF2F}
\BIBentryALTinterwordspacing
K.~Zheng, J.~M. Han, and S.~Polu, ``{MiniF2F}: a cross-system benchmark for
  formal olympiad-level mathematics,'' in \emph{Proc.\ International Conference
  on Learning Representations (ICLR)}, 2022, arXiv:2109.00110. [Online].
  Available: \url{https://arxiv.org/abs/2109.00110}
\BIBentrySTDinterwordspacing

\bibitem{Lample2022HTPS}
G.~Lample, M.-A. Lachaux, T.~Lavril, X.~Martinet, A.~Hayat, G.~Ebner,
  A.~Rodriguez, and T.~Lacroix, ``Hypertree proof search for neural theorem
  proving,'' in \emph{Advances in Neural Information Processing Systems
  (NeurIPS)}, 2022, pp. 27\,959--27\,971.

\bibitem{Trinh2024AlphaGeometry}
T.~H. Trinh, Y.~Wu, Q.~V. Le, H.~He, and T.~Luong, ``Solving olympiad geometry
  without human demonstrations,'' \emph{Nature}, vol. 625, no. 7995, pp.
  476--482, 2024.

\bibitem{DeepMind2025GeminiIMO}
T.~Luong and E.~Lockhart, ``Advanced version of gemini with deep think
  officially achieves gold-medal standard at the international mathematical
  olympiad,''
  \url{https://deepmind.google/discover/blog/advanced-version-of-gemini-with-deep-think-officially-achieves-gold-medal-standard-at-the-international-mathematical-olympiad/},
  Google DeepMind, Jul. 2025, accessed: 2025-07-23.

\bibitem{AppelHaken1977a}
K.~Appel and W.~Haken, ``Every planar map is four colorable. part~i:
  Discharging,'' \emph{Illinois Journal of Mathematics}, vol.~21, no.~3, pp.
  429--490, 1977.

\bibitem{AppelHaken1977b}
K.~Appel, W.~Haken, and J.~Koch, ``Every planar map is four colorable. part~ii:
  Reducibility,'' \emph{Illinois Journal of Mathematics}, vol.~21, no.~3, pp.
  491--567, 1977.

\bibitem{Tucker1977FourColor}
T.~G. Tucker, ``The computer-assisted proof of the four-color theorem,''
  \emph{Scientific American}, vol. 237, no.~4, pp. 128--139, October 1977.

\bibitem{MacKenzie2001Mechanizing}
D.~MacKenzie, \emph{Mechanizing Proof: Computing, Risk, and Trust}.\hskip 1em
  plus 0.5em minus 0.4em\relax MIT Press, 2001.

\bibitem{Buzzard2021Proof}
K.~Buzzard, ``The future of mathematics? formal proof,'' in \emph{The Best
  Writing on Mathematics 2021}, M.~Pitici, Ed.\hskip 1em plus 0.5em minus
  0.4em\relax Princeton University Press, 2021, pp. 121--132.

\bibitem{Heule2016BPT}
M.~J.~H. Heule, O.~Kullmann, and V.~W. Marek, ``Solving and verifying the
  boolean pythagorean triples problem via cube-and-conquer,'' in \emph{Theory
  and Applications of Satisfiability Testing -- SAT 2016}, ser. Lecture Notes
  in Computer Science, vol. 9710.\hskip 1em plus 0.5em minus 0.4em\relax
  Springer, 2016, pp. 228--245.

\bibitem{Heule2018AAAI}
M.~J.~H. Heule, ``Schur number five,'' in \emph{Proceedings of the
  Thirty-Second AAAI Conference on Artificial Intelligence (AAAI-18)}.\hskip
  1em plus 0.5em minus 0.4em\relax AAAI Press, 2018, pp. 6598--6606.

\bibitem{Hales2017Flyspeck}
T.~C. Hales and T.~F. Team, ``A formal proof of the kepler conjecture,''
  \emph{Forum of Mathematics, Pi}, vol.~5, p.~e2, 2017.

\bibitem{Gonthier2013OddOrder}
G.~Gonthier, A.~Mahboubi, and E.~Tassi, ``A small scale reflection extension
  for the coq system,'' in \emph{Interactive Theorem Proving (ITP 2013)}, ser.
  Lecture Notes in Computer Science, vol. 7998.\hskip 1em plus 0.5em minus
  0.4em\relax Springer, 2013, pp. 374--389.

\bibitem{Gonthier2020CFSGSlides}
\BIBentryALTinterwordspacing
G.~Gonthier, ``Toward a formal proof of the classification of finite simple
  groups,'' Slides, Workshop on Formal Mathematics, Fields Institute, 2020.
  [Online]. Available: \url{https://www.mathcomp.org/CFSG-slides-2020}
\BIBentrySTDinterwordspacing

\bibitem{Gentzen1936}
G.~Gentzen, ``Die widerspruchsfreiheit der reinen zahlentheorie,''
  \emph{Mathematische Annalen}, vol. 112, pp. 493--565, 1936.

\bibitem{GorensteinLyonsSolomon1994}
D.~Gorenstein, R.~Lyons, and R.~Solomon, \emph{The Classification of the Finite
  Simple Groups}, ser. Mathematical Surveys and Monographs.\hskip 1em plus
  0.5em minus 0.4em\relax American Mathematical Society, 1994, vol.~40, a
  multi-volume series, starting with Volume 1 in 1994.

\bibitem{Godel1936}
K.~Gödel, ``Über die länge von beweisen,'' \emph{Ergebnisse eines
  Mathematischen Kolloquiums}, vol.~7, pp. 23--24, 1936.

\bibitem{Kreisel1959Proofs}
G.~Kreisel, ``Interpretation of analysis by means of constructive functionals
  of finite types,'' in \emph{Proceedings of the International Congress of
  Mathematicians, 1958}.\hskip 1em plus 0.5em minus 0.4em\relax Cambridge
  University Press, 1959, pp. 255--271.

\bibitem{Kohlenbach2008BK}
U.~Kohlenbach, \emph{Applied Proof Theory: Proof Interpretations and Their Use
  in Mathematics}, ser. Springer Monographs in Mathematics.\hskip 1em plus
  0.5em minus 0.4em\relax Springer, 2008.

\bibitem{HoTTBook2013}
\BIBentryALTinterwordspacing
{The Univalent Foundations Program}, \emph{Homotopy Type Theory: Univalent
  Foundations of Mathematics}.\hskip 1em plus 0.5em minus 0.4em\relax
  {Univalent Foundations Program, Princeton}, 2013, accessed on July 12, 2025.
  [Online]. Available: \url{https://homotopytypetheory.org/book/}
\BIBentrySTDinterwordspacing

\bibitem{deMoura2021Lean}
L.~de~Moura, S.~Ullrich, J.~Avigad, S.~Hudon, and T.~Winterhalter, ``The lean 4
  theorem prover,'' in \emph{Automated Deduction -- CADE 28}, ser. Lecture
  Notes in Computer Science, vol. 12699.\hskip 1em plus 0.5em minus 0.4em\relax
  Springer, 2021, pp. 625--635.

\bibitem{CoqTeam2024}
\BIBentryALTinterwordspacing
{The Coq Development Team}, ``The coq proof assistant (version 8.19) --
  reference manual,'' 2024. [Online]. Available:
  \url{https://coq.inria.fr/refman/}
\BIBentrySTDinterwordspacing

\bibitem{Leroy2009CompCert}
X.~Leroy, ``Formal verification of a realistic compiler,'' \emph{Communications
  of the ACM}, vol.~52, no.~7, pp. 107--115, 2009.

\bibitem{Klein2009seL4}
G.~Klein, K.~Elphinstone, G.~Heiser, J.~Andronick, D.~Cock, P.~Derrin,
  D.~Elkaduwe, K.~Engelhardt, R.~Kolanski, M.~Norrish, T.~Sewell, H.~Tuch, and
  S.~Winwood, ``{seL4}: Formal verification of an {OS} kernel,'' in
  \emph{Proceedings of the 22nd ACM Symposium on Operating Systems Principles
  (SOSP '09)}.\hskip 1em plus 0.5em minus 0.4em\relax ACM, 2009, pp. 207--220.

\bibitem{Kocher2020Spectre}
P.~Kocher, D.~Genkin, D.~Gruss, W.~Haas, M.~Hamburg, M.~Lipp, S.~Mangard,
  T.~Prescher, M.~Schwarz, and Y.~Yarom, ``Spectre attacks: Exploiting
  speculative execution,'' \emph{Communications of the ACM}, vol.~63, no.~7,
  pp. 93--101, 2020.

\bibitem{CoqBug2015}
\BIBentryALTinterwordspacing
M.~D{\'{e}}n{\`{e}}s, ``Coq issue \#4269: A proof of false accepted by
  \texttt{coqchk},'' 2015. [Online]. Available:
  \url{https://github.com/coq/coq/issues/4269}
\BIBentrySTDinterwordspacing

\bibitem{Intel1995FDIV}
M.~J. Flynn and S.~Thibault, ``Intel's pentium floating-point divide fault,''
  \emph{IEEE Micro}, vol.~15, no.~2, pp. 32--40, 1995.

\bibitem{Thompson1984Trust}
K.~Thompson, ``Reflections on trusting trust,'' \emph{Communications of the
  ACM}, vol.~27, no.~8, pp. 761--763, 1984.

\bibitem{Chacon2014ProGit}
\BIBentryALTinterwordspacing
S.~Chacon and B.~Straub, \emph{Pro Git}, 2nd~ed.\hskip 1em plus 0.5em minus
  0.4em\relax Apress, 2014. [Online]. Available:
  \url{https://git-scm.com/book/en/v2}
\BIBentrySTDinterwordspacing

\end{thebibliography}

\end{document}